\documentclass[11pt,twoside]{article}
\usepackage{aom}
\usepackage[english]{babel}

\usepackage[utf8]{inputenc}
\usepackage{tensor}
\usepackage{hyperref}
\let\oldcdot\cdot %if we still need it 
\renewcommand{\cdot}{\mathbin{\mkern-2mu\oldcdot\mkern-2mu}}

\newcommand{\formComma}{\,\text{,}}

\newcommand{\R}{\mathbb{R}}

\newcommand{\surf}{\mathcal{S}}

\newcommand{\paraC}{X}
\newcommand{\para}{\boldsymbol{\paraC}}
\newcommand{\normalC}{\nu}
\newcommand{\normal}{\boldsymbol{\normalC}}

\newcommand{\ofrak}{\mathfrak{o}}

\newcommand{\vnor}{v_{\bot}}

\newcommand{\DX}{D_{\!\para}}

\setAomVolume{N}
\setAomIssue{X}
\setAomPages{pp}
\setAomYear{2026}
\setAomDoi{doi}

\aomAuthorsShort{E. RADICE, I. NITSCHKE, M. SALVALAGLIO, A. VOIGT}
\aomTitleShort{Wrinkling patterns and topological defect dynamics in crystalline films}

\begin{document}

% ---------- Masthead ----------
%\aomMasthead

% ---------- Title / author / affiliation ----------
\title{Wrinkling dynamics of thin crystalline sheets and its dependency on topological defects}

\author{E. RADICE$^{1}$, I. NITSCHKE$^{1}$, M. SALVALAGLIO$^{1,2}$, A. VOIGT$^{1,3,4}$}

% ---------- Affiliations ----------
\address{%
  $^{1}$Institute of Scientific Computing, TU Dresden, Dresden, Germany\\
  $^{2}$Dresden Center for Computational Materials Science, TU Dresden, Germany\\
  $^{3}$Cluster of Excellence -- Physics of Life, TU Dresden, Germany\\
  $^{4}$Center for Systems Biology Dresden (CSBD), Germany\\[4pt]
  %e-mail: emma.radice@tu-dresden.de
}

\begin{aomAbstract}

{\scshape We investigate} the wrinkling dynamics of thin crystalline sheets under uniaxial compression using a Surface Phase-Field-Crystal-Helfrich (SPFCH) model, which captures the coupling between crystalline order, out-of-plane deformations, and topological defects. In particular, the model allows us to track and characterize two distinct classes of defects, namely those forming in the underlying hexagonal crystal lattice, such as dislocations and grain boundaries, and defects in the wrinkling patterns, known as wrinklons. By resolving the evolution and interactions of these defects, we investigate how the interplay between in-plane and out-of-plane relaxation mechanisms influences wrinkling onset and the coarsening regime. Our results indicate that crystalline defects do not affect coarsening scaling, but they quantitatively affect the onset of wrinkling and the emerging wrinkling wavelength, promote more irregular wrinkle patterns with a higher density of wrinklons, and modify long-timescale morphological evolution.
\end{aomAbstract}

\aomKeywords{Wrinkling dynamics, Topological defects, Phase Field Crystal, Thin sheet}

% ============================================================
% ARTICLE BODY
% ============================================================

\section{Introduction}

While thin sheets typically buckle under compression, they also develop wrinkles when attached to a viscous or viscoelastic substrate. For small quasistatic uniaxial compression, classical elasticity theory predicts regular sinusoidal wrinkles with a wavelength determined by the elastic properties of the film as well as the film thickness. Due to their predicted regularity, the resulting equilibrium patterns are appealing for various technical applications, e.g., in light harvesting \cite{Kim2012} or flexible electronics \cite{Khang_2006,Rogers_2010}. However, such wrinkles result from a stress-driven instability with an initial wavelength much shorter than its equilibrium value, and subsequent coarsening. As a consequence of such complex dynamics, small perturbations can critically affect the emergence of regular patterns. We investigate how topological defects in thin crystalline sheets affect the stress-driven instability under uniaxial compression, the coarsening of wrinkles, and, as a result, the regularity of the emerging pattern. 

Understanding the morphological evolution of compressed thin films or sheets and controlling wrinkling broadens their applications in many fields \cite{Chen_2019}. However, even in classical elasticity theory, wrinkling dynamics remain poorly understood. Previous studies predict coarsening laws with a scaling of the wavelength that varies from $t^{1/3}$ \cite{draux2026universal} to
$t^{1/4}$ \cite{ImHuang_2005,huang_dynamics_2006} and $t^{1/6}$ \cite{Vandeparre_2010,Kodio_2017}. This diversity arises from different material properties, computational approaches, and treatments of the substrate's viscous or viscoelastic properties. Because we focus on the impact of topological defects in the wrinkling of thin crystalline sheets, we will follow Ref.~\cite{huang_dynamics_2006} and account for substrate viscosity only implicitly by using a finite mobility coefficient, e.g., an overdamped equation of motion for out-of-plane deformations. Rather than relying on classical elasticity theory, we use the Phase-Field-Crystal (PFC) model \cite{Elder2002,Elder2004}, which resolves crystalline structures on atomic length and diffusive time scales and naturally incorporates crystalline defects and elasticity \cite{Heinonen2014,Huter_nonlinear_2016,Benoit-Marechal_GE_2024}. In particular, we model the thin crystalline sheet using a Surface Phase-Field-Crystal-Helfrich (SPFCH) model, building on \cite{Backofen2010,Backofen2011,BENOITMARECHAL2024105114,radice2026}. We study out-of-plane deformation of a perfect hexagonal crystal, hexagonal crystals hosting an increasing number of isolated dislocations, and a hexagonal
polycrystal with small- and large-angle grain boundaries. We compare the onset of wrinkling in these cases with the theory following from classical elasticity and explore the coarsening of the wrinkle patterns. While averaged quantities for scaling and roughness essentially reproduce the results in \cite{huang_dynamics_2006}, strong deviations appear in the emerging wrinkle patterns and morphological evolution when crystalline defects are considered. These deviations become more pronounced as the number of crystalline defects increases, affecting the regularity of the resulting patterns and also promoting topological defects within the ordered stripe-like patterns that emerge during wrinkling (i.e., wrinklons \cite{vandeparre_wrinkling_2011}). The interplay between crystalline defects, such as dislocations and grain boundaries, and wrinkling patterns under uniaxial compression reveals a complex coupling between atomic-scale processes and macroscopic deformation mechanisms, providing fundamental insights into the mechanical behavior of thin crystalline sheets and potentially enabling new routes to tailor applications.

\section{Methods}

\subsection{Mathematical model}

The Phase-Field-Crystal (PFC) model \cite{Elder2002,Elder2004} operates on atomic length and diffusive time scales and is grounded in dynamical density functional theory \cite{Elder2007,Teeffelen2009}. In its simplest form, it reduces to conserved dynamics (a $H^{-1}$ - gradient flow) of a Swift-Hohenberg free energy ${\mathcal{F}}[\psi]$ with order parameter $\psi$ mimicking the averaged in time atomic density and periodic fields as minimizers representing the crystal lattice. Under small deformations, Hooke's law is recovered \cite{Heinonen2014}, and lattice defects such as dislocations and grain boundaries emerge spontaneously through energy minimization. Thanks to these features, the PFC model enables mesoscale investigation of a broad class of phenomena in crystalline systems \cite{emmerich2012phase,Salvalaglio_2022}, including the
interplay between crystalline order and mechanical deformation as well as defect dynamics \cite{trautt2012coupled,Berry2014,salvalaglio2020coarse,skogvoll2022phase,qiu2024grain}.

Applying this approach to 2D crystals on curved surfaces $\mathcal{S}$ yields the Surface Phase-Field-Crystal (SPFC) model \cite{Backofen2010,Backofen2011,Kohler2016}, which retains all features of the (planar) PFC model except that distances are now measured geodesically.
The model becomes more complex once deformable surfaces are considered. Rather than a straightforward conserved gradient flow acting on $\psi$ alone over a fixed surface, one now needs a gradient flow that minimizes the free energy through the simultaneous evolution of $\psi$ and the surface $\mathcal{S}$. 
We consider a ($L^2, H^{-1}$) - surface gradient flows: a $L^2$ - gradient flow with respect to the evolution of $\mathcal{S}$ and a $H^{-1}$ - gradient flow with respect to the evolution of $\psi$. Besides energy dissipation by evolving $\mathcal{S}$ and $\psi$ simultaneously, we also want to ensure conservation of $\psi$ on the evolving surface, i.e., $\frac{d}{dt} \int_\mathcal{S} \psi d\mathcal{S} = 0$. In order to deal with local expansion or contraction of the surface, the scalar Truesdell time derivative $\mathring{\psi} = \dot{\psi} + \psi\, {\rm Div}_C \mathbf{V}$, where $\dot{\bullet}$ is the material time derivative, ${\rm Div}_C$ is the componentwise trace-divergence, and $\mathbf{V}$ is the material velocity, has been introduced \cite{nitschke2026scalar}. In this way, the Reynolds transport theorem is elegantly incorporated into the mathematical formalism to build ($L^2$, $H^{-1}$) - surface gradient flows that guarantee energy dissipation and conservation properties. Following this general approach, as in \cite{BENOITMARECHAL2024105114,radice2026}, we consider the free energy $\mathcal{F} = \mathcal{F_L} + \mathcal{F_P} + \mathcal{F}_b$ with
\begin{align}
&\mathcal{F_L} = \frac{A}{2} \int_S \left( \psi \mathcal{L}_S\psi\right)d\mathcal{S}, \label{eq:FL} \\
&\mathcal{F_P}=\int_S \left( \frac{B}{2} \psi^2+ \frac{C}{3} \psi^3+\frac{D}{4} \psi^4 \right) d\mathcal{S},\\
&\mathcal{F}_b= \frac{\beta}{2}\int _S \mathcal{H}^2 d\mathcal{S},
\end{align}
with $A, B, C, D$ parameters of the Swift-Hohenberg energy, characterizing the phase space and material properties together with the global average density $\bar{\psi} = \int_{\mathcal{S}} \psi d\mathcal{S}$. The operator $\mathcal{L}_\mathcal{S}= (q^2+\Delta_\mathcal{S} )^2$, with $\Delta_\mathcal{S}$ the Laplace-Beltrami operator, approximates the two-point correlation function and encodes the crystal symmetry, which here corresponds to the hexagonal lattice, i.e., the equilibrium density is well approximated by
\begin{equation}\label{eq:exp}
\psi= \bar{\psi} + \phi_0 \sum_{\mathbf{q}_{\rm hex}} e^{i \mathbf{q}_{\rm {hex}}\cdot \mathbf{r}},
\end{equation}
with $\mathbf{q}_{\rm hex}$ the shortest wavevectors reproducing the hexagonal symmetry (see also Eq.~\eqref{eq:deformation}) \cite{Salvalaglio_2022}:
\begin{equation}\label{eq:qhex}
\{\mathbf{q}_{\rm hex}\}\equiv \left[\pm(0,1),\pm\bigg(\frac{\sqrt{3}}{2},\frac{1}{2}\bigg),\pm\bigg(-\frac{\sqrt{3}}{2},\frac{1}{2}\bigg)\right].
\end{equation}
$\mathcal{F_P}$ addresses the thermotropic part. $\mathcal{F}_b$ is a macroscopic Helfrich energy with $\beta$ the bending rigidity and $\mathcal{H}$ the mean curvature of $\mathcal{S}$. The ($L^2$, $H^{-1}$) - surface gradient flow, which we term Surface Phase-Field-Crystal-Helfrich (SPFCH) model reads 
\begin{align}
    M_{\mathbf{X}} \mathbf{V} &= - \mathcal{D}_{\mathbf{X}} {\mathcal{F}} \label{SGF_1},\\
    M_\psi \mathring{\psi} &= \Delta_{\mathcal{S}} \mathcal{D}_{\psi} {\mathcal{F}}, \label{SGF_2}
\end{align}
where $\mathbf{X}$ denotes a paramaterization of $\mathcal{S}$, $M_\mathbf{X}, M_\psi > 0$ are mobility coefficients and $\mathcal{D}_{\mathbf{X}} {\mathcal{F}}$ and $\mathcal{D}_{\psi} {\mathcal{F}}$ denote functional derivatives leading to $\mathbb{R}^3$ vector fields and scalar fields defined on $\mathcal{S}$. In this form, the surface gradient flow guarantees the conservation property $\frac{d}{dt} \int_\mathcal{S} \psi d\mathcal{S} = 0$ and energy dissipation $\frac{d}{dt} \mathcal{F} \leq 0$. 

In the following, we restrict the surface evolution to the normal direction, e.g., $M_{\mathbf{X}} \vnor = \normal\DX{\mathcal{F}}$, with $\normal$ the outward-pointing normal to $\mathcal{S}$ and $\vnor = \mathbf{V} \cdot \normal$ the normal component of $\mathbf{V}$, and consider a height observer given by the parameterization 
\begin{align}
    \para_{\ofrak}:\ 
        (t,x,y)\mapsto \para_{\ofrak}(t,x,y)
        &:= \left[ x, y, h(t,x,y) \right]^T \in\surf\subset\R^3\formComma
\end{align} 
where $h(t,x,y)$ is the scalar height field and $ (x,y)\in\mathcal{U}\subseteq\R^2 $ are local coordinates. %For many situations, this is sufficient to capture the key phenomenology and has also been considered in \cite{BENOITMARECHAL2024105114,radice2026,PhysRevMaterials.5.034004}. 
The resulting problem corresponding to \eqref{SGF_1} and \eqref{SGF_2} reads:
\begin{align}
\partial_t \psi = &(\partial_t h) (\frac{\langle\nabla h, \nabla \psi\rangle}{|\boldsymbol{g}|} + \psi \Delta_S h ) \nonumber \\
& + M_\psi^{-1} \Delta_S ( A  ( q^4 \psi  + 2q^2 \Delta_S \psi  + \Delta_S^2 \psi )+  B\psi + C \psi^2 + D\psi^3 ), \label{eq_heigt1}\\
\partial_t h = & -M_{\mathbf{X}}^{-1} \sqrt{|\boldsymbol{g}|} (\mathcal{H}\frac{A}{2}((\Delta_S \psi)^2 - q^4\psi^2) - 2 A (\nabla_S \psi \Pi_{QS} \textit{II} \nabla_S (\Delta_S \psi + q^2\psi) ) \nonumber \\
& + \langle\!\langle  A  ( q^4 \psi  + 2q^2 \Delta_S \psi  + \Delta_S^2 \psi ) \rangle\!\rangle  \mathcal{H}\psi+\beta ( 
\Delta_S \mathcal{H} + \mathcal{H} ( \frac{\mathcal{H}^2}{2} - 2\mathcal{K} ) ) ), \label{eq_height2}
\end{align}
with $\boldsymbol{g}$ the metric tensor, $\Pi_{QS} \textit{II}$ the trace-free part of the shape operator, $\mathcal{K}$ the Gaussian curvature (see Appendix \ref{app1}), and $\langle\!\langle\cdot \rangle\!\rangle$ denoting an average over the unit cell via a smoothing kernel in reciprocal space (see Appendix \ref{app:coarse}). Following previous work  \cite{BENOITMARECHAL2024105114,radice2026,PhysRevMaterials.5.034004}, we only considered the elastic parts (neglecting $\mathcal{F_P}$) of $\mathcal{F}$ in \eqref{eq_height2}. This allows for comparison with classical elasticity theories. However, one central difference from such theories, e.g., the Föppl--von Kármán (FvK) equations for thin sheets, remains. The Surface Phase-Field-Crystal-Helfrich (SPFCH) model has dissipative structure. This is expressed by the finite mobility coefficients $M_\psi, M_{\mathbf{X}} > 0$. While $M_\psi$ follows from the PFC modeling approach, a finite $M_{\mathbf{X}}$ essentially realizes viscous dissipation in the surrounding medium, e.g., an underlying viscous or viscoelastic substrate \cite{huang_dynamics_2006}. As a consequence, the model is expected to develop wrinkles under compression without explicitly modeling an underlying dissipative substrate.

\subsection{Stress in the Surface Phase-Field-Crystal (SPFC) model}

The elastic stress field is computed within the Surface Phase-Field-Crystal (SPFC) framework by extending the flat-space formulation of \cite{salvalaglio2019closing,PhysRevB.103.224107} to non-planar surfaces. Elastic deformations of the density entering the PFC model can be characterized by considering the (macroscopic) displacement field $\mathbf{u}$,
\begin{equation}
    \psi(\mathbf{r}, t) = \psi^{\mathrm{ref}}(\mathbf{r} - \mathbf{u}, t) = \bar\psi+\sum_{\mathbf{q}_j}\phi_0 e^{-i\mathbf{q}_j\,\cdot \, \mathbf{u}}e^{i\mathbf{q}_j\,\cdot \, \mathbf{r}},
    \label{eq:deformation}
\end{equation}
where $\psi^{\mathrm{ref}}$ denotes the reference (undistorted) density field ($\mathbf{u}=\mathbf{0}$) and  $\mathbf{q}_j$ the shortest wavevectors reproducing a given lattice symmetry \cite{Salvalaglio_2022}. The macroscopic stress tensor is obtained by spatially coarse-graining the microscopic stress, which is determined by the variation of the free energy functional with respect to mass displacements \cite{salvalaglio2019closing,PhysRevB.103.224107}. We generalize this expression to a curved surface  $\mathcal{S}$ by following a similar approach to \cite{PhysRevA.38.1005,Bassler_biofilm}, i.e., defining a displacement $\mathbf{u} \in  T \mathcal{S}$, with $T \mathcal{S}$ the tangent space of $\mathcal{S}$, replacing flat differential operators with their surface counterparts and thus introducing additional terms through the surface derivatives.
We consider only the stress contributions arising from the elastic part of the energy $\mathcal{F}_{\mathcal{L}}$, so excluding isotropic pressure terms. The microscopic stress tensor for a hexagonal lattice, generalized to a curved surface, reads

\begin{equation}
    \tilde{\sigma}_{ij} = -\left(q^2\psi + \Delta_S\psi\right)\nabla^2_S\psi\big|_{ij} 
    + \nabla_i\!\left(q^2\psi + \Delta_S\psi\right) \, \nabla_j\psi,
\end{equation}
where $i,j \in \{x,y\}$, $\nabla^2_S$ is the Hessian and $\nabla$ is the Euclidean gradient.
We work with the contravariant components by  $\tilde\sigma^{ij}=g^{ik}g^{jl}\tilde\sigma_{kl}$. The stress tensor is intrinsically tangential, $\tilde{\boldsymbol\sigma}\in T\mathcal{S}\otimes T\mathcal{S}$, i.e.\ $\tilde{\boldsymbol\sigma}\,\cdot\,\mathbf{n}=0$, since $\mathbf{u}\in T\mathcal{S}$ by construction. In the Monge parametrization $\mathbf{X}=(x,y,h(x,y))$, the tangent vectors are $\mathbf{e}_x=(1,0,h_x)^T$ and $\mathbf{e}_y=(0,1,h_y)^T$. From this, the stress in the Cartesian components (the stress embedded in  $\mathbb{R}^3$) is computed: 
\begin{equation}
    \tilde{T} = \tilde{\sigma}^{ij}\,\mathbf{e}_i\otimes\mathbf{e}_j = J\,\tilde{\sigma}\,J^T,
    \qquad J=[\mathbf{e}_x\;\mathbf{e}_y].
\end{equation}
$\tilde{T}$ is a microscopic stress. In order to get a macroscopic (hydrodynamic) quantity as in classical continuum mechanics, we average $\tilde{T}$ over the unit cell via a smoothing kernel in reciprocal space (see Appendix \ref{app:coarse}),
\begin{equation}
    T = \langle  \langle \tilde{T} \rangle \rangle  .
\end{equation}
Note that since the Cartesian basis is fixed, the composition of the coarse-graining of each component yields the average of the tensor itself. The coarse-grained tensor is then projected back onto the tangent plane, \(\sigma_{ij}=\mathbf{e}_i^T T\;\mathbf{e}_j\), and normalized by the lengths of the basis vectors, \(\sigma_{\hat i\hat j}=\sigma_{ij}/(|\mathbf{e}_i||\mathbf{e}_j|)\), so that the plotted quantity is a physical stress along the unit vectors $\hat{\mathbf{e}}_i$ and is comparable across regions of different slope. For small out-of-plane deformation, averaging does not significantly alter tangentiality. To discuss the results, we also compute the von Mises stress, a scalar quantity that is independent of the basis in which the stress is expressed. It is defined as \(\sigma_{\rm VM} = \sqrt{3/2 \tau_{ij}\tau_{ij}}\) where $\tau_{ij}$ are the components of the deviatoric stress tensor $T^{\rm dev}= T - \frac{\text{tr}(T)}{3}Id$ with $Id$ the identity in $\mathbb{R}^3$. Since \(\tau_{ij}\tau_{ij}= \text{tr}\;(T ^2) -\frac{1}{3}(\text{tr}\; T)^2\), this yields:  
\[ 
\sigma_{\rm VM} = \sqrt{\frac{3\; \text{tr}\; (T ^2) -(\text{tr}\; T)^2}{2}} .
\]

\subsection{Numerics}
We consider a rectangular domain $(0,L_x)\times(0,L_y)$ with periodic boundary conditions and appropriate initial conditions for $\psi$ and $h$. We solve the model's equation numerically using a Fourier pseudo-spectral method.
The periodic rectangular cell containing two atoms has sides $a_x=4\pi/\sqrt{3}=a$, with $a$ the lattice constant, and $a_y=\sqrt{3}\,a$.
The system size is $L_x=240\,a_x$ and $L_y=160\,a_y$. For discretization, we use a uniform $2048 \times 2048$ grid. The unstrained configuration corresponds to $L_x=256\,a_x$, so that the domain is subject to a compressive strain of $6.25\%$ along $x$ ($\varepsilon_{ij}=-\delta_{ix}\delta_{jx}0.0625$).
 We evaluate the nonlinear terms in physical space and transform to and from Fourier space using the fast Fourier transform (FFT), while we treat the linear terms fully in Fourier space, enabling a time-stepping scheme based on an exponential integrator. For details, see \cite{BENOITMARECHAL2024105114}. The method is implemented on GPU using CuPy, exploiting GPU-accelerated FFT to efficiently handle the large system sizes and long integration times required to resolve both crystalline defect dynamics and wrinkling pattern coarsening. We use a constant time step $\Delta t = t^n - t^{n-1} = 1$ and mobilities
$M_\psi = 1$ and $M_{\mathbf{X}} = 0.1$. $\psi$ is initialized with $\bar\psi=0$ and $\phi_0=(C+\sqrt{C^2-15\;B \;D})/(15\;D)$. We consider $A=1, B=-0.2, C=\frac{ \sqrt{3}}{2\; \sqrt{0.98}}, D=1$, corresponding to previous studies \cite{BENOITMARECHAL2024105114,radice2026}. The bending rigidity is $\beta=1$. 
In the following, lengths, such as the height $h$, are reported in units of $a$.

\subsection{Simulation setup and postprocessing}

\begin{figure}
    \centering
    \includegraphics[width=1\linewidth]{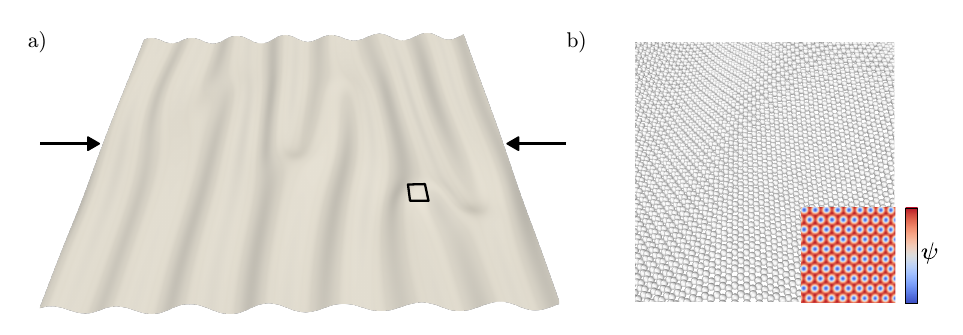}
    \caption{Modeling framework and sample wrinkling morphology. a) Wrinkling of a thin sheet under uniaxial compression ($\varepsilon_{xx} = -6.25 \%$, see black arrows). b)  Hexagonal lattice in the region marked by the black square in panel  a), illustrated by atomic arrangement obtained by reconstructing the peak positions of the density field $\psi$, see inset. Illustration via OVITO \cite{Stukowski_2010}.
    }
    \label{fig:figure0}
\end{figure}

\begin{figure}[t!]
    \centering
    \includegraphics[width=\linewidth]{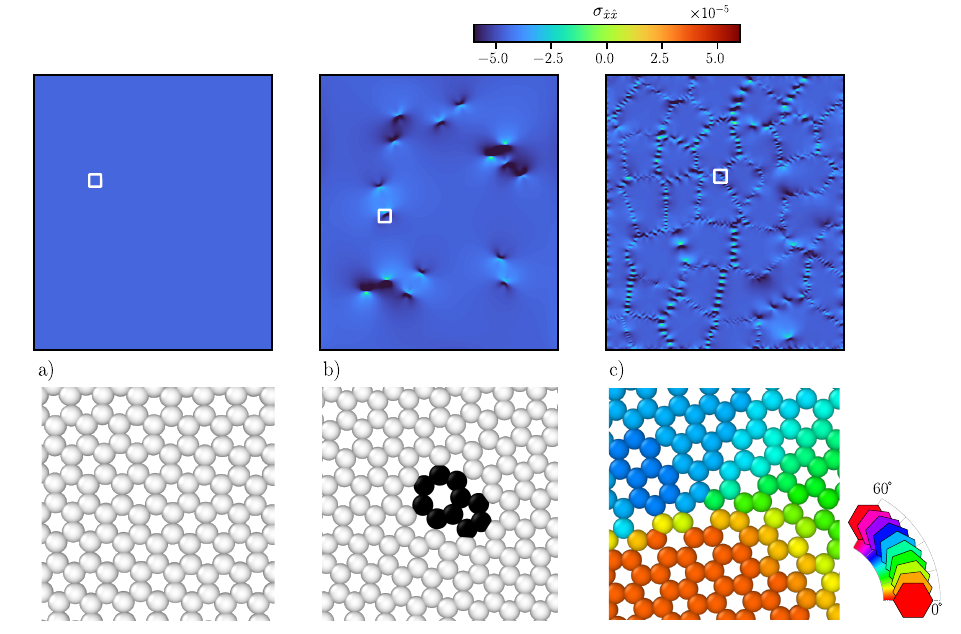}
    \caption{ Stress field projected on $\mathcal{U}$ (first row) and zoomed-in crystalline structure reconstructed as in Figure~\ref{fig:figure0}b) (second row). a) Perfect hexagonal crystal. b) Crystal with $N_\psi=16$ dislocations. The $5-7$ dislocation core is highlighted in black.  c) Polycrystal. Different grain orientations are color-coded.}
    \label{fig:figure1}
\end{figure}

Starting from a flat film, we add a small random perturbation to the initial height field to allow for out-of-plane deformations, breaking the symmetry in which no out-of-plane deformation occurs for a perfectly flat domain ($h(\mathbf{r})=\text{const.}$). We generate the perturbation as Gaussian noise in Fourier space with random amplitudes and phases, imposing zero mean and a fixed rms amplitude of $10^{-5}a$. We choose a small amplitude to avoid biasing the selected wrinkling pattern. We verified with different random seeds that the overall coarsening dynamics are independent of the noise realization, while the positions of the topological defects in the wrinkle arrangement, namely the \textit{wrinklons} \cite{vandeparre_wrinkling_2011}, and the coarsening events differ from run to run. An example of a configuration obtained by solving the model's equations is reported in Figure~\ref{fig:figure0}, illustrating the wrinkled thin sheet and its underlying crystal lattice, along with the field $\psi$ and the reconstructed atomic positions. All simulations use the same initial random height perturbation for direct comparison, but we consider different initial conditions for the crystal lattice, including isolated dislocations and grain boundaries. Figure~\ref{fig:figure1} explicitly illustrates some of these configurations.

The dislocations are inserted by initializing the density via Eq.~\eqref{eq:deformation} and imposing a singularity in the phase $\varphi_i=-\mathbf{q}_i \cdot \mathbf{u}$ \cite{skaugen_dislocation_2017}, such that
\begin{equation}
\oint {\rm d}\varphi_i = \oint {\rm d}\mathbf{u}\cdot \mathbf{q}_i=\mathbf{b}\cdot \mathbf{q}_i=2\pi s,
\end{equation}
for a $\mathbf{q}_i\in \{\mathbf{q}_{\rm hex}\}$, see Eq.~\eqref{eq:qhex}, with $s$ the winding number and $\mathbf{b}$ the dislocation Burgers vector, thus linked to the lattice symmetry via the scalar product with $\mathbf{q}_i$. This can be realized by imposing $\varphi_j(\mathbf{r})=s (2 \pi)^{-1} \theta (\mathbf{r}-\mathbf{r}_0)$ for a given $j$, with $\mathbf{r}_0$ the nominal dislocation core location and $s=1$, i.e., introducing a perfect dislocation. We always introduce dislocations as dipoles, i.e., pairs with opposite Burgers vectors, to ensure zero total topological charge while still imposing random positions. We impose a minimum distance between all cores to avoid overlaps in the initial condition. The lattice deformation induced by dislocations is then obtained after a first relaxation of the initial condition. We recall that the resulting elastic field agrees with continuum-mechanics predictions \cite{skaugen_dislocation_2017,skogvoll2022phase,BENOITMARECHAL2024105114,Benoit-Marechal_GE_2024}.

We initialize polycrystalline films by placing $N_{\rm seeds}=40$ hexagonal crystal seeds with randomly assigned orientations in the range $[0,\pi/3)$. Seed positions are randomly sampled within the computational domain, subject to a minimum separation of ($d_{\min}=0.8\min(L_x,L_y)/\sqrt{N_{\rm seeds}}$) from all previously accepted seeds. Each grain is represented by a rotated hexagonal lattice modulated by a Gaussian envelope with width ($\min(L_x,L_y)/(2\sqrt{N_{\rm seeds}})$). During subsequent relaxation, the grains grow and eventually impinge on one another, forming grain boundaries.

The uniaxial compression induces out-of-plane deformations, creating a corrugated profile. Wrinkles orient to relax the deformation (decrease the film's elastic energy), forming aligned stripes perpendicular to the compression axis. Because of the initial Gaussian noise, in the early stages of pattern formation the stripe alignment is not yet fully established and the pattern can display local irregularities including topological defects, i.e., wrinklons \cite{vandeparre_wrinkling_2011}. 
We locate them by demodulating the height field around the dominant wavevector $\mathbf{k}_0$ (see Appendix \ref{app:plots}), namely computing $z=|z|e^{i\zeta}$ such that $h-\bar{h}=ze^{i\mathbf{k}_0\,\cdot\, \mathbf{r}}+{\rm c.c.}$ with $\bar{h}$ the average height, and counting its phase windings~\cite {skogvoll_unified}.
$$w=\frac{1}{2\pi}\oint d\zeta\;\in\;\mathbb Z.$$
The sign of $w$ gives the wrinklon's topological charge. 

The analysis of the wrinkling dynamics is performed by computing a few representative quantities and studying their evolution over time. They are: the mean wavelength of the wrinkled height profile $\bar{\lambda}(t)$, the root-mean-square amplitude of the surface profile $W_{\rm RMS}(t)$ (the roughness), the root-mean-square Gaussian curvature of the surface $K_{\rm RMS}$, the mean von Mises stress $\bar\sigma_{\rm VM}$, the number of crystalline defects $N_\psi(t)$ and the number of wrinklons $N_{h}(t)$. Details concerning their evaluation are reported in the Appendix~\ref{app:plots}. 

\section{Results}

  We consider uniaxial compression of a thin sheet,
  as illustrated in Figure \ref{fig:figure0}. We first derive an analytic result, instrumental for discussing the numerical results of the full Surface Phase-Field-Crystal-Helfrich (SPFCH) model. Then, we illustrate and analyze the morphological evolution and deformation of thin crystalline sheets hosting different defect configurations. By comparing them, we derive general trends on the impact of crystalline defects on wrinkling dynamics.

\subsection{Derivation of the onset wrinkling wavelength}

To describe the out-of-plane deformation instability within the one-dimensional continuum framework, we consider the 1D Föppl--von Kármán (FvK) equations for a thin film, which we couple to an overdamped dynamics for the height field $h$:
\begin{subequations}
\label{eq:fvk_system}
\begin{align}
0 &= \frac{\mathrm{d}}{\mathrm{d}y} \left[ \frac{\mathrm{d}u}{\mathrm{d}y} + \frac{1}{2} \left( \frac{\mathrm{d}h}{\mathrm{d}y} \right)^2 \right], \label{eq:fvk_inplane} \\
\frac{1}{\Gamma}\frac{\partial h}{\partial t} &= -\beta \frac{\partial^4 h}{\partial y^4} + (\lambda + 2\mu) \frac{\mathrm{d}}{\mathrm{d}y} \left[ \left( \frac{\mathrm{d}u}{\mathrm{d}y} + \frac{1}{2} \left( \frac{\mathrm{d}h}{\mathrm{d}y} \right)^2 \right) \frac{\mathrm{d}h}{\mathrm{d}y} \right], \label{eq:fvk_outplane}
\end{align}
\end{subequations}
where $u(y)$ is the in-plane displacement, $h(y,t)$ represents the out-of-plane deflection (height), $\beta$ is the bending rigidity, $1/\Gamma$ is the mobility coefficient, equivalent to $M_{\mathbf{X}}$ and containing the viscosity of the substrate, and $\lambda, \mu$ are the Lamé elastic parameters. Following the same computation as \cite{huang_dynamics_2006}, under a uniform in-plane compressive (negative) strain  $\varepsilon_0 < 0$,  we have $ \frac{\mathrm{d}u}{\mathrm{d}y} + \frac{1}{2} ( \frac{\mathrm{d}h}{\mathrm{d}y} )^2 = -|\varepsilon_0|$ and the onset wavelength can be determined. Substituting this relation into Eq.~\eqref{eq:fvk_outplane} yields the linearized overdamped equation of motion for small out-of-plane perturbations:
\begin{equation}
\frac{\partial h}{\partial t} = -\Gamma \left[ \beta \frac{\partial^4 h}{\partial y^4} + (\lambda + 2\mu) |\varepsilon_0| \frac{\partial^2 h}{\partial y^2} \right].
\label{eq:linearized_h}
\end{equation}
By assuming a plane-wave perturbation of the form $h(y, t) = \hat{h} \, e^{\gamma(k) t} e^{i k y}$, with $\gamma(k)$ growth rate, and inserting in \eqref{eq:linearized_h} we obtain:
\begin{equation}
\gamma(k) = \Gamma \left[ (\lambda + 2\mu) |\varepsilon_0| k^2 - \beta k^4 \right].
\label{eq:dispersion}
\end{equation}
The dominant mode $k^*$ corresponding to the maximum growth rate ($\mathrm{d}\gamma / \mathrm{d}k = 0$) is given by:
\begin{equation}
k^* = \sqrt{\frac{(\lambda + 2\mu) |\varepsilon_0|}{2\beta}},
\label{eq:k_star}
\end{equation}
which defines the characteristic onset wavelength:
\begin{equation}
\lambda_{\text{onset}} = \frac{2\pi}{k^*} = 2\pi \sqrt{\frac{2\beta}{(\lambda + 2\mu) |\varepsilon_0|}}.
\label{eq:lambda_fvk}
\end{equation}
Without energy dissipation in the substrate, accounted for by the viscosity in the mobility coefficient $1/\Gamma$, e.g. for $\Gamma \to 0$, the result drastically changes and we obtain the typical buckling of thin sheets under compression \cite{Seung1988}.

Within the PFC model framework, and assuming small deformation (linear elasticity), the effective in-plane elastic modulus for a defect-free hexagonal crystal is $(\lambda + 2\mu) = 9 A \phi_0^2$. $A$ is the coefficient in \eqref{eq:FL} and $\phi_0$ is the (real) equilibrium amplitude of the lattice in \eqref{eq:exp}. Nonlinearities lead to a (small) correction to the elastic modulus of $\left(1- \frac{5}{4}\varepsilon_0\right)$ as discussed in \cite{BENOITMARECHAL2024105114}.   Substituting this elasticity relation into Eq.~\eqref{eq:lambda_fvk} yields the explicit onset wavelength as a function of the simulation parameters:
\begin{equation}
\lambda_{\text{onset}} = 2\pi \sqrt{\frac{2\beta}{9A \phi_0^2 \left(1-\frac{5}{4}\varepsilon_0\right)  \, |\varepsilon_0|}  } \approx 8.36.
\label{eq:lambda_sapfc}
\end{equation}
This provides a simple validation criterion for the numerical simulations based on the Surface Phase-Field-Crystal-Helfrich (SPFCH) model.

\subsection{Numerical Simulations}

Besides the examples in Figure \ref{fig:figure1}, namely the sheets with $N_\psi = 0$ and $N_\psi = 16$ dislocations, as well as a polycrystalline sheet, we also consider $N_\psi= 8, 60, 100$ to inspect various distributions and degrees of plastic relaxation. The polycrystalline case serves as a representative case, featuring both small- and large-angle grain boundaries. 

\begin{figure}
    \centering
    \includegraphics[width=1\linewidth]{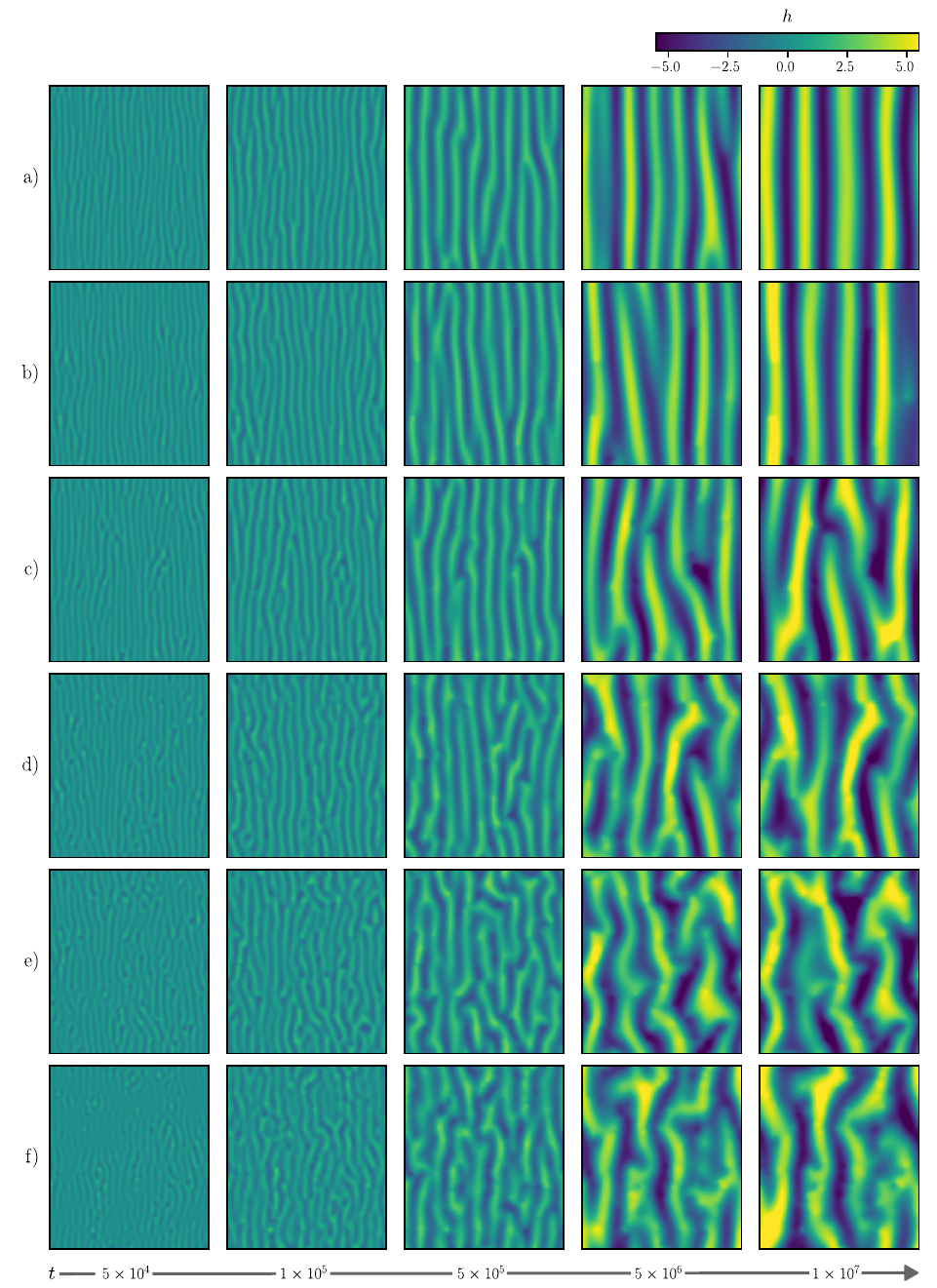}
    \caption{Height profile $h$ at representative stages: $t= 5 \times 10 ^4, 1  \times10^5, 5 \times 10^5, 5 \times 10^6, 1 \times 10^7$. a) Perfect crystal, b)-e) Crystal with isolated dislocations with $N_\psi = 8, 16, 60, 100$, respectively. f) Polycrystal.}
    \label{fig:Height profiles}
\end{figure}

\begin{figure}
    \centering
    \includegraphics[width=1\linewidth]{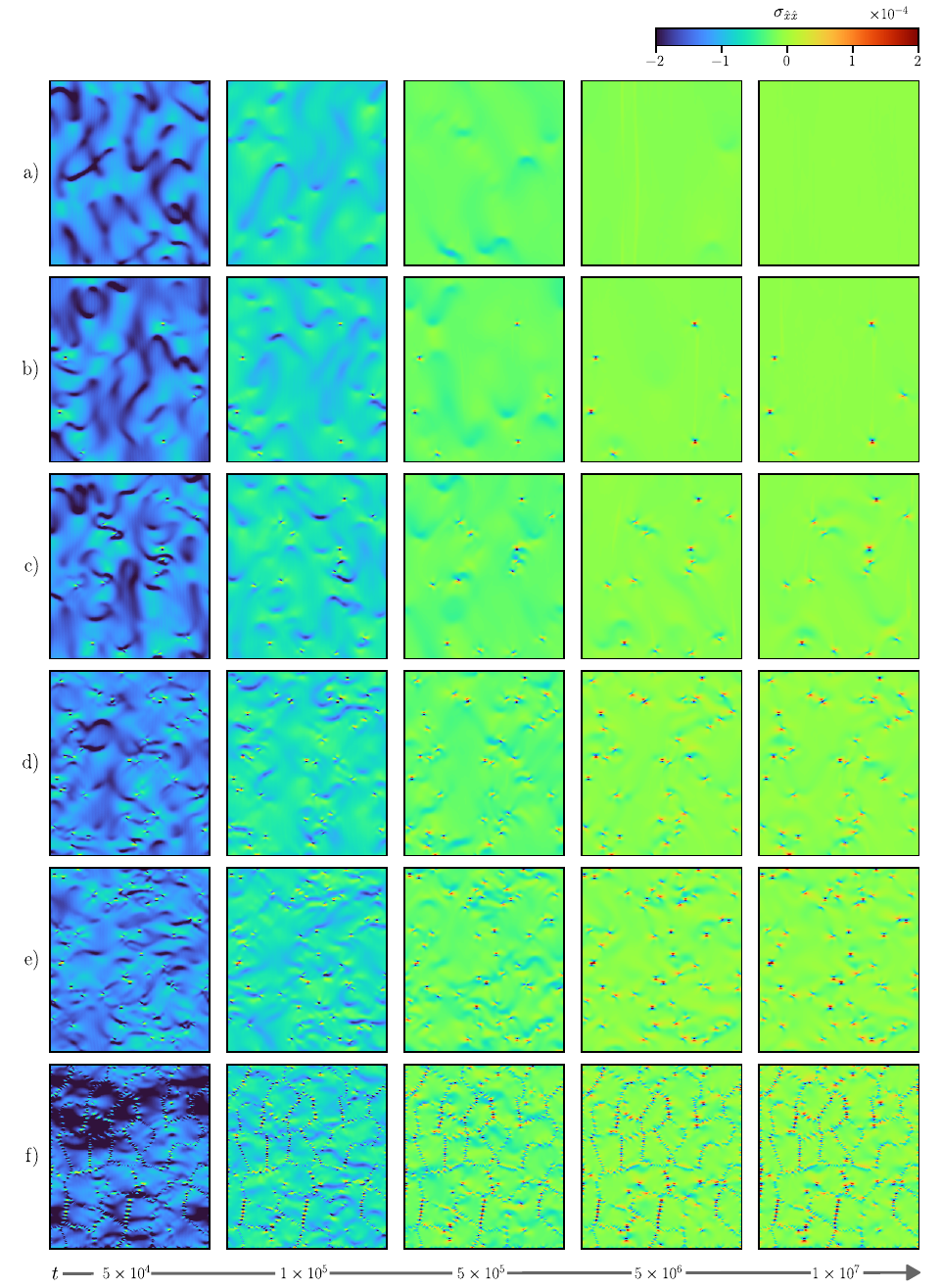}
    \caption{Stress field $\sigma_{xx}$ at representative stages: $t= 5 \times 10 ^4, 1  \times10^5, 5 \times 10^5, 5 \times 10^6, 1 \times 10^7$. a) Perfect crystal, b)-e) Crystal with isolated dislocations with $N_\psi = 8, 16, 60, 100$, respectively. f) Polycrystal.}
    \label{fig:stressprofiles}
\end{figure}

The considered examples show that different distributions and types of crystalline defects (dislocations and grain boundaries) introduce different deformation fields and affect wrinkling patterns and their dynamics differently.
Figure~\ref{fig:Height profiles} illustrates the morphological evolution of thin sheets hosting such distributions over long times through representative stages. The wrinkling instability is associated with the development of a pattern featuring wrinklons both in the presence and absence of initial crystalline defects. Without dislocations or grain boundaries, a regular pattern develops, eventually leading after coarsening to the buckled state with a wavelength equal to the domain size $L_x$ (not shown).
As defect density increases, we observe the nucleation of more wrinklons, while the emerging height-profile pattern becomes more irregular. With high crystalline dislocation density, the wrinkle crests also strongly deviate from directions perpendicular to the uniaxial deformation, with persistent irregularities. Figure \ref{fig:stressprofiles} illustrates the evolution of thin-sheet stress field $\sigma_{xx}$ at the same stages shown in Figure~\ref{fig:Height profiles}. In all cases, the initial compression relaxes. Higher defect density leads to larger stress variations in the crystal between dislocations, which in turn increases the wrinklon density.

\begin{figure}[t!]
    \centering
    \includegraphics[width=1\linewidth]{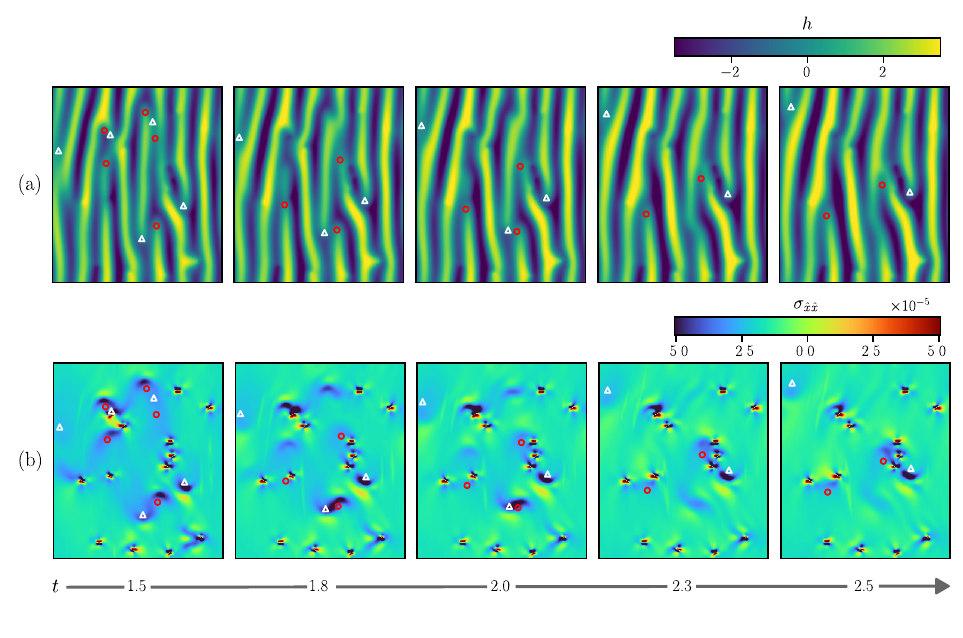}
    \caption{Wrinkling dynamics for the case with $N_\psi = 16$ initial dislocations. Representative stages are selected. a) Height profile $h$, wrinklons with $w=+1$ are marked with red circles and with $w=-1$ with white triangles. b) Stress field of configurations in a) projected on $\mathcal{U}$ together with winklons positions.}
    \label{fig:figure2}
\end{figure}

While Figures~\ref{fig:Height profiles} and \ref{fig:stressprofiles} clearly indicate the increased irregularity in the height field and its relation to crystalline defects, they do not reveal the microscopic mechanisms mediating complex morphological evolution. To better understand them, Figure~\ref{fig:figure2} reports a small intermediate time sequence of the case with initially $N_\psi=16$ dislocations, corresponding to Figure \ref{fig:Height profiles} b) and Figure \ref{fig:stressprofiles} b), focusing on the annihilation of wrinklons. Wrinklons with positive and negative charges ($w=+1$ red circles $w=-1$ white triangles) are illustrated. The total topological charge $\sum_i w_i=0$ vanishes at each time instance, and the number of wrinklons decreases over time due to these annihilation events. The evolution of the height profile reveals a progressive coarsening of the wrinkling pattern, driven by the migration and subsequent annihilation of wrinklon pairs, as illustrated in Figure  \ref{fig:figure2}. The wrinklons are also identifiable in the corresponding stress profiles, where they give rise to pronounced localized stress variations (Figure  \ref{fig:figure2}b). These regions constrain out-of-plane deformation, hindering stress relaxation by forming an ideal wrinkling pattern. 

\begin{figure}[t!]
    \centering
    \includegraphics[width=1\linewidth]{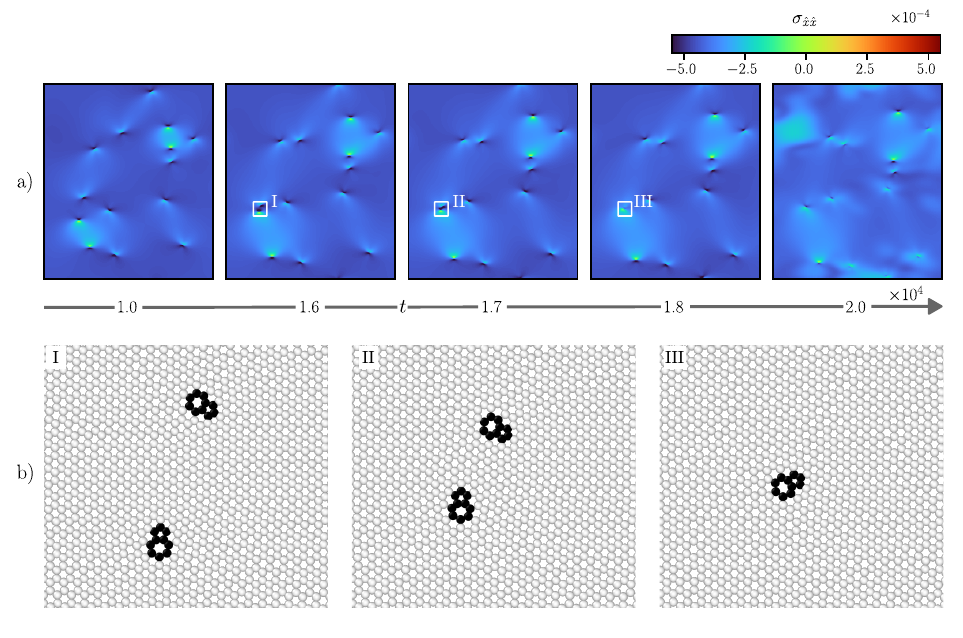}
    \caption{Example of dislocation reaction from the case with $N_\psi = 16$ initial dislocations. Representative stages are selected. a) Stress field projected on $\mathcal{U}$. b) Hexagonal lattice in the white squares of panel a) and dislocation cores illustrated as in Figure~\ref{fig:figure1}b).}
    \label{fig:figure3}
\end{figure}

As evident from Figure \ref{fig:figure2}, crystalline defects such as dislocations also migrate and interact as the system evolves. They also cause localized perturbations in the height profiles \cite{BENOITMARECHAL2024105114,radice2026,dedonno2026}. In Figure \ref{fig:figure3}, an early intermediate time sequence of the case with initially $N_\psi=16$ dislocations, corresponding to Figure \ref{fig:stressprofiles} b), is reported. We concentrate on the interaction of two dislocations. In the case illustrated there, they approach and react, forming a new dislocation with a different orientation. Annihilation also occurs when two dislocations with equal and opposite Burgers vectors react. We do not report the height field at the dislocation cores, as the out-of-plane deformation is negligible compared to the wrinkling. Detailed investigations of out-of-plane deformations of isolated dislocations using the Surface Phase-Field-Crystal-Helfrich (SPFCH) model are available in \cite{radice2026}. 

\begin{figure}
    \centering
    \includegraphics[width=1\linewidth]{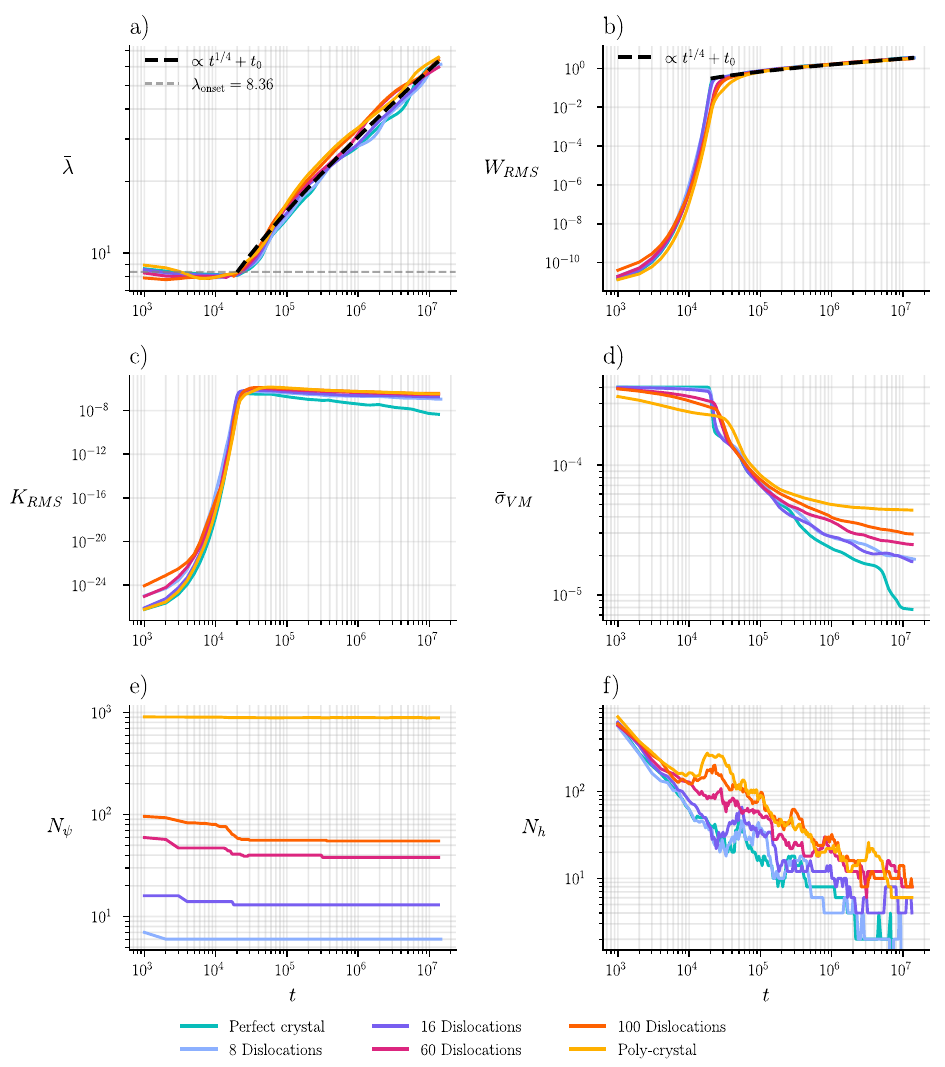}
    \caption{Analysis of the coarsening dynamics. a) Average wavelength $\bar{\lambda}(t)$, together with theoretical  wavelength from Eq. \eqref{eq:lambda_sapfc} and $t^{1/4}$ power law. 
    b) Root-mean-square amplitude $W_{\rm RMS}(t)$, together with $t^{1/4}$ power law.  
    c) Root mean square Gaussian curvature $K_{\rm RMS}(t)$. d) Mean von Mises stress $\bar{\sigma}_{\rm VM}(t)$. e) Number of crystalline defects $N(t)$. f) Number of wrinklons, detected with winding number $\pm 1$, $N_{h}(t)$. The cases correspond to the wrinkling dynamics reported in Figure \ref{fig:Height profiles} and Figure \ref{fig:stressprofiles}.}

    \label{fig:plots}
\end{figure}

The dynamics illustrated above show that the interplay of these processes, namely the migration and annihilation of wrinklons as well as the migration and annihilation/reaction of crystalline defects, affects the wrinkling dynamics. In order to quantify and characterize related effects, we compute a few representative quantities (detailed in the Appendix \ref{app:plots}) and study their evolution over time, see Figure~\ref{fig:plots}.

For the defect-free sheet, the onset wavelength for wrinkling agrees with the theoretical prediction in Eq.~\eqref{eq:lambda_sapfc}. With crystalline defects, the same process occurs, but we observe minor deviations from this wavelength. Furthermore, after an initial transient, the wavelength coarsening dynamics follow the theoretically predicted $t^{1/4}$ scaling \cite{huang_dynamics_2006} across all considered systems, regardless of the density or nature of the crystalline defects. This behavior is illustrated by the black dashed line in Figure~\ref{fig:plots} a). These findings indicate that crystalline defects influence the initial wrinkling wavelength but do not significantly alter its power-law scaling over time.

The surface roughness $W_{\rm RMS}$ likewise exhibits $t^{1/4}$ power-law scaling across all considered systems, consistent with \cite{huang_dynamics_2006}. This behavior is illustrated in Figure~\ref{fig:plots} b), where the black dashed line represents this scaling. Notably, although the scaling exponent remains unchanged, the onset of this regime is progressively delayed with increasing defect density, with the most pronounced delay observed in the polycrystalline system. These results suggest that crystallographic defects partially hinder wrinkle development by delaying amplitude growth, without altering the underlying scaling behavior observed above.

The root-mean-square Gaussian curvature $K_{\rm RMS}$ in Figure~\ref{fig:plots} c) exhibits a qualitatively similar evolution across all considered systems during the initial growth of the wrinkle amplitudes, with only minor differences. At later stages, however, this quantity decreases markedly in the defect-free crystal, whereas it decreases much less in the presence of crystalline defects. In the latter cases, the decrease is at least one order of magnitude smaller than in the defect-free system and becomes progressively more pronounced as the defect density decreases.

Stress and topological defects evolve differently across all considered cases. The mean von Mises stress, $\bar{\sigma}_{\rm VM}$, is shown in Figure \ref{fig:plots} d). It initially decreases, with a more pronounced relaxation observed at higher defect densities. This behavior can be attributed to defect rearrangements and the associated plastic flow, which are absent in the perfect crystal. 

The initial relaxation is also accompanied by defect annihilation, as shown in Figure \ref{fig:plots} e). As wrinkling enters the coarsening regime, out-of-plane deformation becomes the dominant relaxation mechanism, leading to a marked change in the rate of stress reduction. The stress change in this phase is more pronounced at lower defect densities, indicating more efficient out-of-plane deformation in those cases. At later stages, when the wrinkles reach larger wavelengths, the remaining crystalline defects increasingly determine the residual stress. Consequently, $\bar{\sigma}_{\rm VM}$ increases with defect density, consistent with the dislocation populations shown in Figure \ref{fig:plots} e). 

A particularly notable difference appears in the evolution of the number of wrinklons, shown in Figure \ref{fig:plots} f). This number increases with dislocation density and reaches its highest value in the polycrystalline system. Crystallographic defects are then found to trigger these defects in the height profile. This also explains the mechanism behind the delays in the growth of the wrinkled profile amplitude and related associated stress release: see the number of wrinklons and the change in the slope of $\bar{\sigma}_{\rm VM}(t)$ and $W_{\rm RMS}(t)$. Note also that, at late times, the drop of $\bar{\sigma}_{\rm VM}(t)$ for the case of a perfect crystal without dislocations can be associated with the formation of perfectly aligned wrinkles (no wrinklons), see Figure \ref{fig:plots} f).

\section{Conclusions}

Wrinkling dynamics of thin sheets under uniaxial compression has previously been explored by classical elasticity theory, identifying two regimes:
a stress-driven instability with an initial wrinkle wavelength much shorter than the equilibrium value, corresponding to buckling of the entire film, and wrinkle coarsening until they reach an equilibrium wavelength. Averaged quantities like the initial wavelength and the scaling exponent for the coarsening process depend on the considered material, but seem to be universal with respect to the strength of uniaxial compression upon rescaling. Here, we explored how emerging wrinkling patterns and their coarsening depend on the material's crystalline microstructure, particularly on the presence of topological defects of different kinds.

Thanks to the Surface Phase-Field-Crystal-Helfrich (SPFCH model, we can capture the coupling between in-plane and out-of-plane relaxation mechanisms. This coupling is central to the physics at play, as it allows the system to redistribute imposed compressive stress between two competing channels: in-plane relaxation, mediated by  defect migration/annihilation and grain boundary motion within the crystalline lattice, and out-of-plane relaxation, mediated by bending and the emergence of a wrinkling pattern. A key strength of the model is its ability to simultaneously investigate two distinct classes of topological defects and their interplay: dislocations and grain boundaries in the crystalline order, described by $\psi$, and wrinklons in the out-of-plane wrinkling pattern, described by $h$.

Our results show that the emergence and evolution of the wrinkling pattern are influenced by the underlying microstructure of the crystal. Crystalline defects are found to slightly affect the wrinkling onset wavelength and, more significantly, the growth of the emerging wrinkles. Still, they play a negligible role in the scaling law for the increase in wrinkle wavelength during coarsening over time. Higher crystalline defect density leads to the nucleation of more wrinklons.
This is also associated with more irregular and persistent wrinkling patterns and less efficient stress relaxation at later stages. These findings highlight that crystalline defects primarily control the morphology and relaxation efficiency of wrinkled thin sheets rather than their underlying coarsening laws.

We remark that our results are consistent with, and to some extent reproduce, predictions of classical elasticity theory for wrinkling and coarsening \cite{ImHuang_2005,huang_dynamics_2006}. However, for various applications such as strain sensors, functional surfaces, flexible electronics, and energy storage  \cite{Chen_2019}, the observed differences in wrinkling patterns, the presence of wrinklons, and the overall long-term morphological evolution may play a crucial role in determining the mechanical and functional properties of thin films. These aspects are found to be crucially coupled with crystalline microstructure, with additional phenomenology that can be relevant for optimization and novel design concepts. Extending these findings to specific materials and such applications would require material-specific parametrizations and calibration against experimental conditions. Nevertheless, the general mechanisms and trends identified in this work provide a foundation for in-depth understanding and potentially controlling the morphological evolution of crystalline thin films.

% ---------- Acknowledgments (optional) ----------
\subsection*{Acknowledgments}
 E.R. and A.V. acknowledge support from DurAMat under the European Union - Marie Sk\l{}odowska-Curie Actions Doctoral Networks (ITN) Call H2020-MSCA-DN-2022, grant no. 101119767. M.S. acknowledges support from the Deutsche Forschungsgemeinschaft (DFG, German Research Foundation), project numbers 417223351 (FOR3013). The authors gratefully acknowledge the computing time made available to them on the high-performance computer at the NHR Center of TU Dresden. This center is jointly supported by the Federal Ministry of Research, Technology and Space of Germany and the state governments participating in the \href{www.nhr-verein.de/unsere-partner}{NHR}.

\section{Appendix}  

\subsection{Height formalism}  \label{app1}

 In order to formulate the equations in local coordinates $(x,y)\in\mathcal{U}\subseteq\R^2$ use:   
%The induced metric, its determinant and its inverse are
\[
  g_{ij}=\delta_{ij}+\partial_i h \; \partial_j h,\qquad
  |\mathbf{g}|=\det(g_{ij})=1+|\nabla h|^2,\qquad
  g^{ij}=\delta^{ij}-\frac{\partial_i h \; \partial_j h}{|\mathbf{g}|},
\]

\[
  \mathcal{H} = \frac{1}{\sqrt{|\mathbf{g}|}}\left( \Delta h
  - \frac{\langle \nabla h, (\nabla^2 h)\,\nabla h \rangle}{|\mathbf{g}|} \right), \qquad
  K = \frac{\det(\nabla^2 h)}{|\mathbf{g}|^2},
\]
\[
  \langle \nabla_S f, \nabla_S \ell \rangle_S
  = \langle \nabla f , \nabla \ell \rangle -
  \frac{\langle \nabla h, \nabla f \rangle \langle \nabla h, \nabla \ell \rangle}{|\mathbf{g}|},
\]
\[
  \Delta_S f 
  = \Delta f
  - \frac{\langle \nabla h, \nabla f \rangle}{|\mathbf{g}|} \Delta h
  - \frac{\langle \nabla h, (\nabla^2 f)\,\nabla h \rangle}{|\mathbf{g}|}
  + \frac{\langle \nabla h, (\nabla^2 h)\,\nabla h \rangle \langle \nabla h, \nabla f \rangle}{|\mathbf{g}|^2}.
\]
\[
  \nabla_S f = \nabla f-\frac{\langle \nabla h,\nabla f\rangle}{|\mathbf{g}|}\,\nabla h ,\qquad
  \nabla_S^2 f
  =  \nabla^2 f-\frac{\langle \nabla h,\nabla f\rangle}{|\mathbf{g}|}\, \nabla^2 h .
\]

\[
  \textit{II} = \frac{ \nabla^2 h}{\sqrt{|\mathbf{g}|}}, \qquad
  (\Pi_{QS}\textit{II}) = \textit{II} - \frac{\mathcal{H}}{2}\,\mathrm{Id}_{TS}
\]
 For $f, \ell : \mathcal{U} \to \R$ and $\mathrm{Id}_{TS}$ is the identity operator in the tangential space,  $\sqrt{|\mathbf{g}|}$ is the area element. Here $\nabla$, $\nabla^2$, $\Delta$ and
$\langle\cdot,\cdot\rangle$ (no subscript $S$) are the Euclidean operators in $(x,y)$.

\subsection{Quantitative measures of wrinkling evolution}\label{app:plots}
Let $h_{ij}$ denote $h$ on the $N_x\times N_y$ grid, $\bar h=\frac{1}{N_xN_y}\sum_{i,j}h_{ij}$,
and $\hat h(\mathbf k)$ the discrete Fourier transform of $h-\bar h$. In Figure \ref{fig:plots} are reported  the roughness, $W_{\rm RMS}(t)$, the power-weighted RMS wavelength $\bar\lambda(t)$ in agreement with \cite{huang_dynamics_2006}.
\begin{align}
  W_{\rm RMS}(t) &= \sqrt{\frac{1}{N_xN_y}\sum_{i,j}\left(h_{ij}-\bar h\right)^2}, \\
  \bar k(t) &= \sqrt{\frac{\sum_{\mathbf k}|\hat h(\mathbf k)|^2\,|\mathbf k|^2}
                          {\sum_{\mathbf k}|\hat h(\mathbf k)|^2}},
  \qquad \bar\lambda(t)=\frac{2\pi}{\bar k(t)},
\end{align}

Gaussian curvature and stress are also reported. Since they are scalar fields on $\mathcal S$, they are averaged with respect to the surface measure $d\mathcal{A}=\sqrt{|\mathbf g|}\,dx\,dy=\sqrt{1+|\nabla h|^2}\,dx\,dy$, with $\mathcal{A}=\int_{\mathcal S}d\mathcal{A}$:
\begin{equation}
  K_{\rm RMS}=\sqrt{\frac{1}{\mathcal{A}}\int_{\mathcal S}K^2\,d\mathcal{A}},
  \qquad
  \bar\sigma_{\rm VM}=\frac{1}{\mathcal{A}}\int_{\mathcal S}\sigma_{\rm VM}\,d\mathcal{A} .
\end{equation}

To compute the number of wrinklons, we start by writing $h- \bar h$ as a slowly modulated plane wave, $(h- \bar h) (\mathbf{r})\simeq z (\mathbf{r})\,e^{i\mathbf k_0\cdot\mathbf r}+\mathrm{c.c.}$, where $\mathbf{k_0}$ is the dominant peak of the pattern spectrum. $\mathbf{k_0}$ is shifted to the origin of Fourier space, and a Gaussian filter of width $|\mathbf{k_0}|$ suppresses the conjugate band at $-2\mathbf{k_0}$. 
$$z=\mathcal F^{-1}\!\Big[\mathcal F\big[(h-\bar h)\,e^{-i\mathbf k_0\cdot\mathbf r}\big]\,e^{-|\mathbf k|^2/2(|\mathbf k_0|)^2}\Big],$$
with  $F\big[f(\mathbf{r})\big]=\hat f\mathbf{(k)}$ and $F^{-1}\!\big[ \hat f\mathbf{(k)} \big]=f(\mathbf{r)}$. The modulus of $z$, $|z|$, is the local pattern amplitude and the argument, $\zeta=\arg z$, is the local phase. At a dislocation core, $|z|\to0$ and the phase winds by a multiple of $2\pi$. Following a similar approach as \cite{Happel26}, we sum the wrapped phase differences $\mathrm{wrap}(\Delta\zeta)\in(-\pi,\pi]$ around each elementary $2\times2$ pixels plaquette of the grid (with periodic boundaries),
$$w=\frac{1}{2\pi}\oint d\zeta\in\mathbb Z,$$
and flag every plaquette with $w\neq0$ as a defect, with the sign of $w$ giving its topological charge.
\\Defects in the density field $\psi$ are identified through a Delaunay triangulation of the local minima of $\psi$, each of which corresponds to a lattice cell (the maxima being the lattice sites). In a perfect lattice, every minimum has coordination number $6$ (six nearest neighbours), while dislocations are detected as bound $5 - 7$ pairs. Periodic boundary conditions are taken into account.

\subsection{Coarse-graining}\label{app:coarse}
The coarse-graining procedure is obtained by averaging a microscopic quantity $\tilde{f}$ over the unit cell with a Gaussian smoothing kernel in reciprocal space to get its macroscopic counterpart $f$
\begin{equation}
    f = \langle\langle \tilde{f} \rangle\rangle
      = \mathcal{F}^{-1}\!\left[\hat{G}(\mathbf{k})\,\mathcal{F}[\tilde{f}](\mathbf{k})\right],
    \qquad
    \hat{G}(\mathbf{k}) = \exp\!\left[-\tfrac{1}{2}\left(a_x^2 k_x^2 + a_y^2 k_y^2\right)\right],
\end{equation}
where $a_x=4\pi/\sqrt{3}$ and $a_y=4\pi$.
% ============================================================
% BIBLIOGRAPHY (BibTeX, references.bib)
% ============================================================
\bibliographystyle{unsrt}
\bibliography{references}

@article{BENOITMARECHAL2024105114,
title = {Mesoscale modeling of deformations and defects in thin crystalline sheets},
journal = {Mechanics of Materials},
volume = {198},
pages = {105114},
year = {2024},
issn = {0167-6636},
doi = {https://doi.org/10.1016/j.mechmat.2024.105114},
url ={https://www.sciencedirect.com/science/article/pii/S0167663624002060},
author = {Lucas Benoit–Maréchal and Ingo Nitschke and Axel Voigt and Marco Salvalaglio}}

@article{Benoit-Marechal_GE_2024,
doi = {10.1088/1361-651X/ad42bb},
year = {2024},
publisher = {IOP Publishing},
volume = {32},
number = {5},
pages = {055005},
author = {Lucas Benoit-Maréchal and Marco Salvalaglio},
title = {{Gradient elasticity in Swift–Hohenberg and phase-field crystal models}},
journal = {Model. Simul. Mater. Sci. Eng.}
}

@article{Huter_nonlinear_2016,
  title = {Nonlinear elastic effects in phase field crystal and amplitude equations: Comparison to ab initio simulations of bcc metals and graphene},
  author = {H{\"u}ter, Claas and Fri{\'a}k, Martin and Weikamp, Marc and Neugebauer, J{\"o}rg and Goldenfeld, Nigel and Svendsen, Bob and Spatschek, Robert},
  journal = {Phys. Rev. B},
  volume = {93},
  issue = {21},
  pages = {214105},
  numpages = {14},
  year = {2016},
  publisher = {American Physical Society},
  doi = {10.1103/PhysRevB.93.214105},
  url = {https://link.aps.org/doi/10.1103/PhysRevB.93.214105}
}

@article{trautt2012coupled,
  title={{Coupled motion of asymmetrical tilt grain boundaries: Molecular dynamics and phase field crystal simulations}},
  author={Trautt, ZT and Adland, A and Karma, A and Mishin, Y},
  journal={Acta Mater.},
  volume={60},
  number={19},
  pages={6528--6546},
  year={2012},
  publisher={Elsevier},
  doi = {10.1016/j.actamat.2012.08.018}
}

@article{qiu2024grain,
  title={Grain boundaries are Brownian ratchets},
  author={Qiu, Caihao and Punke, Maik and Tian, Yuan and Han, Ying and Wang, Siqi and Su, Yishi and Salvalaglio, Marco and Pan, Xiaoqing and Srolovitz, David J and Han, Jian},
  journal={Science},
  volume={385},
  number={6712},
  pages={980--985},
  year={2024},
  publisher={American Association for the Advancement of Science},
  doi = {10.1126/science.adp1516}
}

@article{skogvoll2022phase,
  title={A phase field crystal theory of the kinematics of dislocation lines},
  author={Skogvoll, Vidar and Angheluta, Luiza and Skaugen, Audun and Salvalaglio, Marco and Vi{\~n}als, Jorge},
  journal={J. Mech. Phys. Solids},
  volume={166},
  pages={104932},
  year={2022},
  publisher={Elsevier},
  doi = "10.1016/j.jmps.2022.104932"
}

@article{Berry2014,
  title = {Phase field crystal modeling as a unified atomistic approach to defect dynamics},
  author = {Berry, Joel and Provatas, Nikolas and Rottler, J\"org and Sinclair, Chad W.},
  journal = {Phys. Rev. B},
  volume = {89},
  pages = {214117},
  numpages = {13},
  year = {2014},
  publisher = {American Physical Society},
  doi = {10.1103/PhysRevB.89.214117}
}

@article{PhysRevB.103.224107,
  title = {Stress in ordered systems: Ginzburg-Landau-type density field theory},
  author = {Skogvoll, Vidar and Skaugen, Audun and Angheluta, Luiza},
  journal = {Physical Review B},
  volume = {103},
  issue = {22},
  pages = {224107},
  numpages = {18},
  year = {2021},
  publisher = {American Physical Society},
  doi = {10.1103/PhysRevB.103.224107},
  url = {https://link.aps.org/doi/10.1103/PhysRevB.103.224107}
}

@article{salvalaglio2019closing,
  title={Closing the gap between atomic-scale lattice deformations and continuum elasticity},
  author={Salvalaglio, Marco and Voigt, Axel and Elder, Ken R},
  journal={npj Computational Materials},
  volume={5},
  number={1},
  pages={48},
  year={2019},
  publisher={Nature Publishing Group UK London}
}

@article{salvalaglio2020coarse,
  title={A coarse-grained phase-field crystal model of plastic motion},
  author={Salvalaglio, Marco and Angheluta, Luiza and Huang, Zhi-Feng and Voigt, Axel and Elder, Ken R and Vi{\~n}als, Jorge},
  journal={Journal of the Mechanics and Physics of Solids},
  volume={137},
  pages={103856},
  year={2020},
  publisher={Elsevier}
}

@article{Elder2002,
  title = {Modeling {{Elasticity}} in {{Crystal Growth}}},
  author = {Elder, K. R. and Katakowski, Mark and Haataja, Mikko and Grant, Martin},
  year = {2002},
  journal = {Physical Review Letters},
  volume = {88},
  number = {24},
  pages = {245701},
  publisher = {{American Physical Society}},
  doi = {10.1103/PhysRevLett.88.245701},
  urldate = {2023-03-08}
}

@article{Elder2004,
  title = {Modeling Elastic and Plastic Deformations in Nonequilibrium Processing Using Phase Field Crystals},
  author = {Elder, K. R. and Grant, Martin},
  year = {2004},
  journal = {Physical Review E},
  volume = {70},
  number = {5},
  pages = {051605},
  publisher = {{American Physical Society}},
  doi = {10.1103/PhysRevE.70.051605},
  urldate = {2022-01-07}
}

@article{Elder2007,
  title = {Phase-Field Crystal Modeling and Classical Density Functional Theory of Freezing},
  author = {Elder, K. R. and Provatas, Nikolas and Berry, Joel and Stefanovic, Peter and Grant, Martin},
  year = {2007},
  journal = {Physical Review B},
  volume = {75},
  number = {6},
  pages = {064107},
  publisher = {{American Physical Society}},
  doi = {10.1103/PhysRevB.75.064107},
  urldate = {2022-01-07}
}

@article{Teeffelen2009,
  title={Derivation of the phase-field-crystal model for colloidal solidification},
  author={van Teeffelen, Sven and Backofen, Rainer and Voigt, Axel and L{\"o}wen, Hartmut},
  journal={Physical Review E},
  volume={79},
  doi = {10.1103/PhysRevE.79.051404},
  number={5},
  pages={051404},
  year={2009},
  publisher={APS}
}

@article{Heinonen2014,
  title = {Phase-Field-Crystal Models and Mechanical Equilibrium},
  author = {Heinonen, V. and Achim, C. V. and Elder, K. R. and Buyukdagli, S. and {Ala-Nissila}, T.},
  year = {2014},
  journal = {Physical Review E},
  volume = {89},
  number = {3},
  pages = {032411},
  publisher = {{American Physical Society}},
  doi = {10.1103/PhysRevE.89.032411},
  urldate = {2023-03-08}
}

@article{PhysRevMaterials.5.034004,
  title = {Modeling buckling and topological defects in stacked two-dimensional layers of graphene and hexagonal boron nitride},
  author = {Elder, K. R. and Achim, C. V. and Heinonen, V. and Granato, E. and Ying, S. C. and Ala-Nissila, T.},
  journal = {Physical Review Materials},
  volume = {5},
  issue = {3},
  pages = {034004},
  numpages = {21},
  year = {2021},
  publisher = {American Physical Society},
  doi = {10.1103/PhysRevMaterials.5.034004},
  url = {https://link.aps.org/doi/10.1103/PhysRevMaterials.5.034004}
}

@article{Backofen2010,
  title = {Particles on curved surfaces: A dynamic approach by a phase-field-crystal model},
  author = {Backofen, Rainer and Voigt, Axel and Witkowski, Thomas},
  journal = {Physical Review E},
  volume = {81},
  issue = {2},
  pages = {025701},
  numpages = {4},
  year = {2010},
  publisher = {American Physical Society},
  doi = {10.1103/PhysRevE.81.025701}
}

@article{Kohler2016,
  title = {Stress {{Induced Branching}} of {{Growing Crystals}} on {{Curved Surfaces}}},
  author = {K{\"o}hler, Christian and Backofen, Rainer and Voigt, Axel},
  year = {2016},
  journal = {Physical Review Letters},
  volume = {116},
  number = {13},
  pages = {135502},
  publisher = {{American Physical Society}},
  doi = {10.1103/PhysRevLett.116.135502},
  urldate = {2023-03-31}
}

@article{Seung1988,
  title = {Defects in Flexible Membranes with Crystalline Order},
  author = {Seung, H. S. and Nelson, David R.},
  year = {1988},
  journal = {Physical Review A},
  volume = {38},
  number = {2},
  pages = {1005--1018},
  publisher = {{American Physical Society}},
  doi = {10.1103/PhysRevA.38.1005},
  urldate = {2023-03-31}
}

@article{Backofen2011,
  title = {A {{Continuous Approach}} to {{Discrete Ordering}} on $\mathbb{S}^2$},
  author = {Backofen, Rainer and Gr{\"a}f, Manuel and Potts, Daniel and Praetorius, Simon and Voigt, Axel and Witkowski, Thomas},
  year = {2011},
  journal = {Multiscale Modeling \& Simulation},
  volume = {9},
  number = {1},
  pages = {314--334},
  publisher = {{Society for Industrial and Applied Mathematics}},
  issn = {1540-3459},
  doi = {10.1137/100787532},
  urldate = {2023-03-31}
}

@article{nitschke2026scalar,
  title={Scalar {T}ruesdell time derivative and {$(L^2$,$H^{-1})$}-surface gradient flows},
  author={Nitschke, Ingo and Voigt, Axel},
  journal={arXiv:2604.08186},
  year={2026}
}

@article{Stukowski_2010,
doi = {10.1088/0965-0393/18/1/015012},
url = {https://doi.org/10.1088/0965-0393/18/1/015012},
year = {2009},
publisher = {},
volume = {18},
number = {1},
pages = {015012},
author = {Stukowski, Alexander},
title = {Visualization and analysis of atomistic simulation data with OVITO–the Open Visualization Tool},
journal = {Modelling and Simulation in Materials Science and Engineering},
}

@article{PhysRevA.38.1005,
  title = {Defects in flexible membranes with crystalline order},
  author = {Seung, H. S. and Nelson, David R.},
  journal = {Phys. Rev. A},
  volume = {38},
  issue = {2},
  pages = {1005--1018},
  numpages = {0},
  year = {1988},
  publisher = {American Physical Society},
  doi = {10.1103/PhysRevA.38.1005},
  url = {https://link.aps.org/doi/10.1103/PhysRevA.38.1005}
}

@article{
Bassler_biofilm,
author = {Chenyi Fei  and Sheng Mao  and Jing Yan  and Ricard Alert  and Howard A. Stone  and Bonnie L. Bassler  and Ned S. Wingreen  and Andrej Košmrlj },
title = {Nonuniform growth and surface friction determine bacterial biofilm morphology on soft substrates},
journal = {Proceedings of the National Academy of Sciences},
volume = {117},
number = {14},
pages = {7622-7632},
year = {2020},
doi = {10.1073/pnas.1919607117},
URL = {https://www.pnas.org/doi/abs/10.1073/pnas.1919607117},
eprint = {https://www.pnas.org/doi/pdf/10.1073/pnas.1919607117},
}

@article{huang_dynamics_2006,
	title = {Dynamics of wrinkle growth and coarsening in stressed thin films},
	volume = {74},
	copyright = {http://link.aps.org/licenses/aps-default-license},
	issn = {1539-3755, 1550-2376},
	url = {https://link.aps.org/doi/10.1103/PhysRevE.74.026214},
	doi = {10.1103/PhysRevE.74.026214},
	language = {en},
	number = {2},
	urldate = {2026-08-17},
	journal = {Physical Review E},
	author = {Huang, Rui and Im, Se Hyuk},
	year = {2006},
	pages = {026214},
}

@article{vandeparre_wrinkling_2011,
	title = {Wrinkling {Hierarchy} in {Constrained} {Thin} {Sheets} from {Suspended} {Graphene} to {Curtains}},
	volume = {106},
	copyright = {http://link.aps.org/licenses/aps-default-license},
	issn = {0031-9007, 1079-7114},
	url = {https://link.aps.org/doi/10.1103/PhysRevLett.106.224301},
	doi = {10.1103/PhysRevLett.106.224301},
	language = {en},
	number = {22},
	urldate = {2026-08-17},
	journal = {Physical Review Letters},
	author = {Vandeparre, Hugues and Piñeirua, Miguel and Brau, Fabian and Roman, Benoit and Bico, José and Gay, Cyprien and Bao, Wenzhong and Lau, Chun Ning and Reis, Pedro M. and Damman, Pascal},
	year = {2011},
	pages = {224301},
}

@article{skaugen_dislocation_2017,
  title = {Dislocation dynamics and crystal plasticity in the phase-field crystal model},
  author = {Skaugen, Audun and Angheluta, Luiza and Vi{\~n}als, Jorge},
  journal = {Phys. Rev. B},
  volume = {97},
  issue = {5},
  pages = {054113},
  numpages = {10},
  year = {2018},
  publisher = {American Physical Society},
  doi = {10.1103/PhysRevB.97.054113},
  url = {https://link.aps.org/doi/10.1103/PhysRevB.97.054113}
}

@article{Salvalaglio_2022,
doi = {10.1088/1361-651X/ac681e},
url = {https://doi.org/10.1088/1361-651X/ac681e},
year = {2022},
publisher = {IOP Publishing},
volume = {30},
number = {5},
pages = {053001},
author = {Salvalaglio, Marco and Elder, Ken R},
title = {Coarse-grained modeling of crystals by the amplitude expansion of the phase-field crystal model: an overview},
journal = {Modelling and Simulation in Materials Science and Engineering},
}

@article{radice2026,
      title={Surface Phase-Field-Crystal-Helfrich model for out-of-plane deformations in thin crystalline sheets with lattice mismatch}, 
      author={Emma Radice and Ingo Nitschke and Marco Salvalaglio and Axel Voigt},
      year={2026},
      journal={arXiv:2607.12997},
      url={https://arxiv.org/abs/2607.12997}, 
}

@article{emmerich2012phase,
  title={Phase-field-crystal models for condensed matter dynamics on atomic length and diffusive time scales: an overview},
  author={Emmerich, Heike and L{\"o}wen, Hartmut and Wittkowski, Raphael and Gruhn, Thomas and T{\'o}th, Gyula I and Tegze, Gy{\"o}rgy and Gr{\'a}n{\'a}sy, L{\'a}szl{\'o}},
  journal={Advances in Physics},
  volume={61},
  number={6},
  pages={665--743},
  year={2012},
  publisher={Taylor \& Francis}
}

@article{draux2026universal,
  title={Universal wrinkling dynamics of a sheet on viscous liquid},
  author={Draux, Ayrton and Rizzo, Marco and Vella, Dominic and D{\'e}mery, Vincent and Brau, Fabian and Damman, Pascal},
  journal={arXiv:2601.11682},
  year={2026}
}

@article{ImHuang_2005,
    author = {Im, S. H. and Huang, R.},
    title = {Evolution of Wrinkles in Elastic-Viscoelastic Bilayer Thin Films},
    journal = {Journal of Applied Mechanics},
    volume = {72},
    number = {6},
    pages = {955-961},
    year = {2005},
    doi = {10.1115/1.2043191},
    url = {https://doi.org/10.1115/1.2043191},
}

@article{Vandeparre_2010,
    author = {Vandeparre, Hugues and Gabriele, Sylvain and Brau, Fabian and Gay, Cyprien and Parker, Kevin Kit and Damman, Pascal},
    title = {Hierarchical wrinkling patterns},
    journal = {Soft Matter},
    volume = {6},
    number = {22},
    pages = {5751-5756},
    year = {2010},
    doi = {10.1039/c0sm00394h},
    url = {https://doi.org/10.1039/c0sm00394h},
}

@article{Kodio_2017,
  title = {Lubricated wrinkles: Imposed constraints affect the dynamics of wrinkle coarsening},
  author = {Kodio, Ousmane and Griffiths, Ian M. and Vella, Dominic},
  journal = {Phys. Rev. Fluids},
  volume = {2},
  issue = {1},
  pages = {014202},
  numpages = {19},
  year = {2017},
  publisher = {American Physical Society},
  doi = {10.1103/PhysRevFluids.2.014202},
  url = {https://link.aps.org/doi/10.1103/PhysRevFluids.2.014202}
}

@article{Chen_2019,
    author = {Chen, Wenjun and Gui, Xuchun and Yang, Leilei and Zhu, Hai and Tang, Zikang},
    title = {Wrinkling of two-dimensional materials: methods, properties and applications},
    journal = {Nanoscale Horizons},
    volume = {4},
    number = {2},
    pages = {291-320},
    year = {2019},
    doi = {10.1039/c8nh00112j},
    url = {https://doi.org/10.1039/c8nh00112j},
}

@article{Khang_2006,
author = {Dahl-Young Khang  and Hanqing Jiang  and Young Huang  and John A. Rogers },
title = {A Stretchable Form of Single-Crystal Silicon for High-Performance Electronics on Rubber Substrates},
journal = {Science},
volume = {311},
number = {5758},
pages = {208-212},
year = {2006},
doi = {10.1126/science.1121401},
URL = {https://www.science.org/doi/abs/10.1126/science.1121401},
}

@article{Rogers_2010,
author = {John A. Rogers  and Takao Someya  and Yonggang Huang },
title = {Materials and Mechanics for Stretchable Electronics},
journal = {Science},
volume = {327},
number = {5973},
pages = {1603-1607},
year = {2010},
doi = {10.1126/science.1182383},
URL = {https://www.science.org/doi/abs/10.1126/science.1182383},
}

@article{Kim2012,
  author    = {Kim, Jong Bok and Kim, Pilnam and P{\'e}gard, Nicolas C. and Oh, Soong Ju and Kagan, Cherie R. and Fleischer, Jason W. and Stone, Howard A. and Loo, Yueh-Lin},
  title     = {Wrinkles and deep folds as photonic structures in photovoltaics},
  journal   = {Nature Photonics},
  year      = {2012},
  volume    = {6},
  number    = {5},
  pages     = {327--332},
  doi       = {10.1038/nphoton.2012.70},
  url       = {https://doi.org/10.1038/nphoton.2012.70},
  issn      = {1749-4893}
}

@article{Happel26,
    author = {Happel, Lea and Oberschelp, Griseldis and Richter, Anneli and Rode, Gwenda Roselene and Grudtsyna, Valeriia and Doostmohammadi, Amin\
 and Voigt, Axel},
    title = {A spectrum of p-atic symmetries and defects in confluent epithelia},
    journal = {Soft Matter},
    volume = {22},
    number = {5},
    pages = {1109-1129},
    year = {2026},
    issn = {1744-683X},
    doi = {10.1039/d5sm01010a},
    url = {https://doi.org/10.1039/d5sm01010a}
}

@article{skogvoll_unified,
    author = {Skogvoll, Vidar and Rønning, Jonas and Salvalaglio, Marco and Angheluta, Luiza },
    title = {A unified field theory of topological defects and non-linear local excitations},
    journal = {npj Computational Materials} ,
    volume= {9},
    article={122},
    year = {2023}
}

@article{dedonno2026,
    author = {De Donno, Marcello and Angheluta, Luiza and Salvalaglio, Marco},
    title = {Dislocation dynamics on deformable surfaces},
    journal = {Proc. Royal Soc. A},
    volume = {482},
    number = {2347},
    pages = {20260141},
    year = {2026},
    month = {10},
    issn = {1364-5021},
    doi = {10.1098/rspa.2026.0141}
}

\end{document}